\documentstyle[prl,aps,epsf,twocolumn]{revtex}
\begin{document}

\draft

\title{
Resonantly enhanced nonlinear optics in semiconductor quantum wells:
An application to sensitive infrared detection 
}
\author{S.~F.~Yelin$^{1,2}$ and P.~R.~Hemmer$^2$}
\address{
        $^1$ ITAMP, Harvard-Smithsonian Center for Astrophysics,
		Cambridge, MA~02138 \\
	$^2$ Sensor's Directorate, Hanscom Air Force Research Laboratory, Hanscom, MA 01731 }
\date{\today}
\maketitle

\begin{abstract}

A novel class of coherent nonlinear optical phenomena, involving
induced transparency 
in semiconductor quantum wells, is considered in the context of a particular application
to sensitive long-wavelength infrared detection. It is shown that the
strongest decoherence mechanisms can be suppressed or mitigated,
resulting in substantial enhancement of nonlinear optical effects in
semiconductor quantum wells.

\end{abstract}

\pacs{PACS numbers 42.50.Gy, 42.65.-k, 78.67.De, 85.60.Gz}


Theoretical and experimental work of the past few years has led to   
a renaissance in the field of resonant nonlinear optics
\cite{EIT}. This work is   
based on quantum coherence and interference effects such as
electromagnetically induced transparency (EIT). Under certain 
conditions they allow to eliminate the resonant absorption and control
the refractive index, and simultaneously  enhance nonlinearities. 
 
For example studies involving second harmonic generation
\cite{hakuta91}, phase conjugation \cite{hemmer95}, 
nonlinear spectroscopy \cite{lukin97}, and coherent Raman scattering 
\cite{hakuta97} promise to improve considerably the performance of novel
nonlinear optical mechanisms.

In the present Letter we show that these improvements can be used to
make resonantly enhanced nonlinear optics feasible 
in semiconductor quantum well systems. Coherence based
nonlinear optics eliminates the need for phase matching and strong
fields. Being able to incorporate these novel methods into
semiconductor materials would be a basis for small and practical devices
utilizing nonlinear optics in engineerable structures with desirable
properties and wavelengths.

As a specific example of such nonlinear phenomena in
semiconductor quantum wells we concentrate on a coherence based variety of
quantum well infrared photo detector (QWIP). Here the presence of
infrared (IR) radiation can modify the transmission spectrum for light
of an easier-to-access 
wavelength. In our example the two fields are strongly coupled via
resonant tunneling \cite{faist,sera00}. 

In particular, we address the challenges connected with decoherence,
which, in semiconductor nanostructures, 
is a much more demanding problem than in more simple, e.g. atomic,
systems. Thus we anticipate that the present
approach can be also useful in a number of other applications such as
efficient switching and modulation.

We begin by illustrating the basic principle of coherence based photo
detection, using 
a generic four-state system. For the moment we assume that all coherent couplings in the
scheme are accomplished by external monochromatic electromagnetic radiation.

\begin{figure}[ht]
\epsfxsize=6cm
\centerline{\epsffile{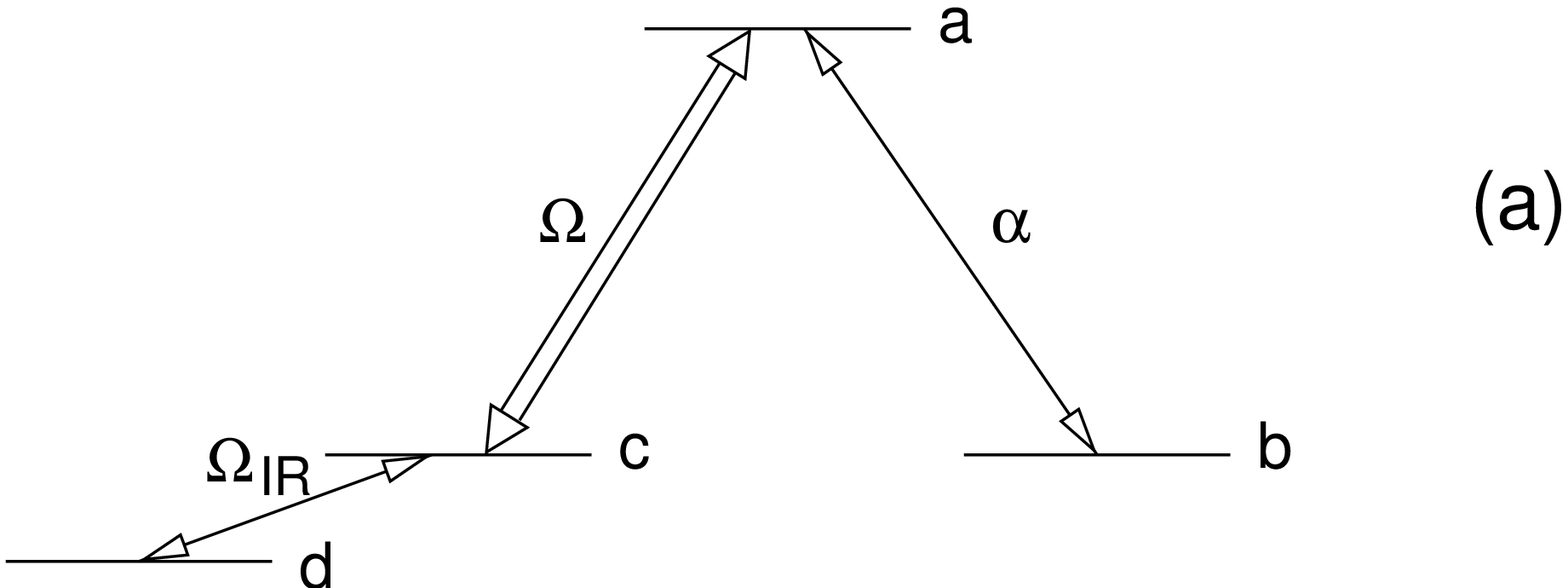}}
\epsfxsize=6cm
\centerline{\epsffile{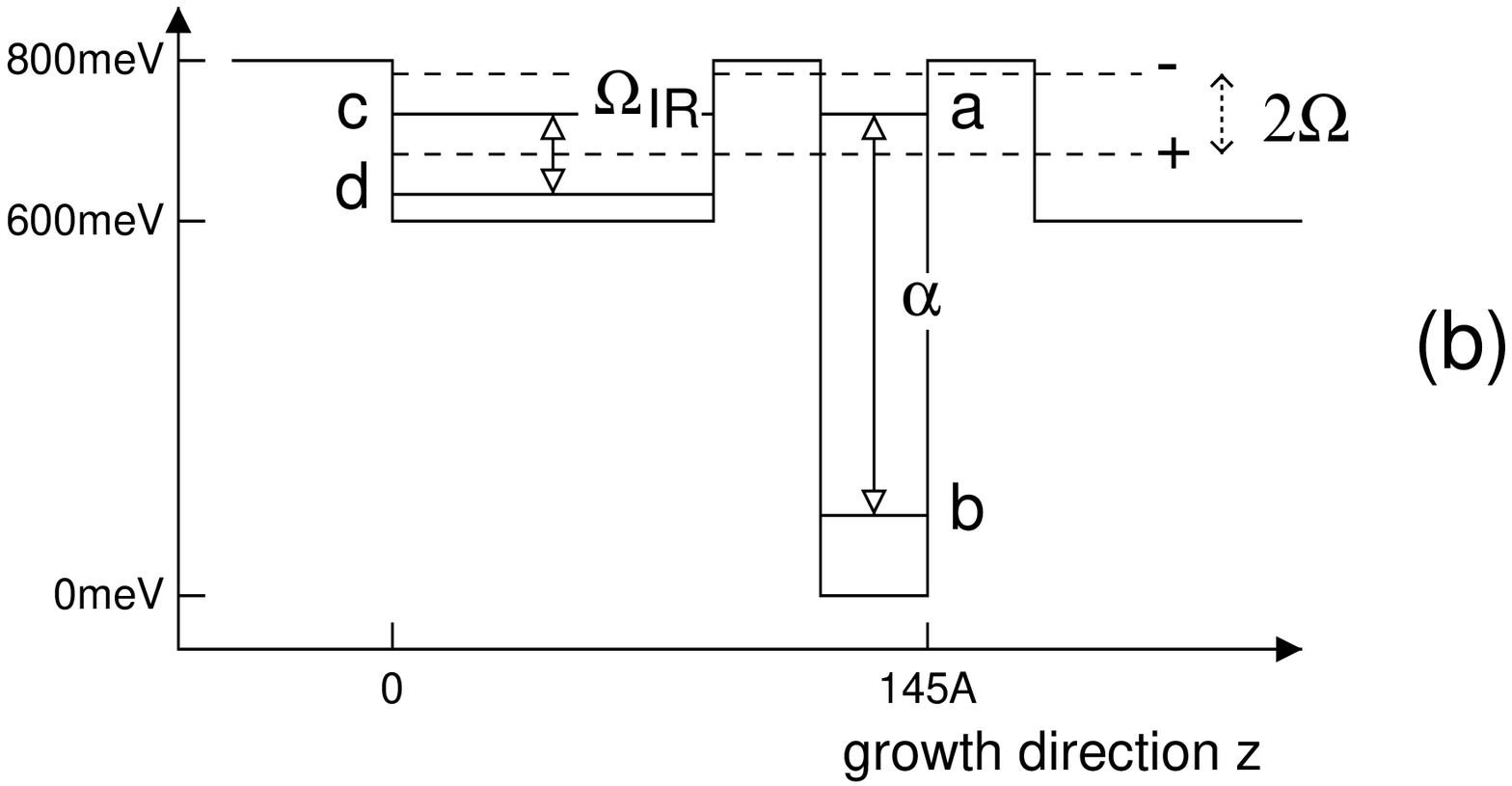}}
\caption{\protect\label{f_levels}
 (a) $\Lambda$-system with states $\left| a \right\rangle$, $\left| b \right\rangle$, and $\left| c \right\rangle$, and fields $\Omega$
and $\alpha$. Additional weak coupling $\Omega_{\rm IR}$ to state $\left|
b \right\rangle$.
 (b) Same system but in a double well. Uncoupled well states $\left| a \right\rangle$ and
$\left| c \right\rangle$ are 
connected by resonant tunneling. The dashed levels
$\left| +\right\rangle\sim(\left| a \right\rangle+\left| c \right\rangle)/\sqrt{2}$ and $\left| -\right\rangle\sim(\left| a
\right\rangle-\left| c \right\rangle)/\sqrt{2}$ are the
eigenstates in this double well system.}
\end{figure}

\begin{figure}[ht]
\epsfxsize=5.5cm
\centerline{\epsffile{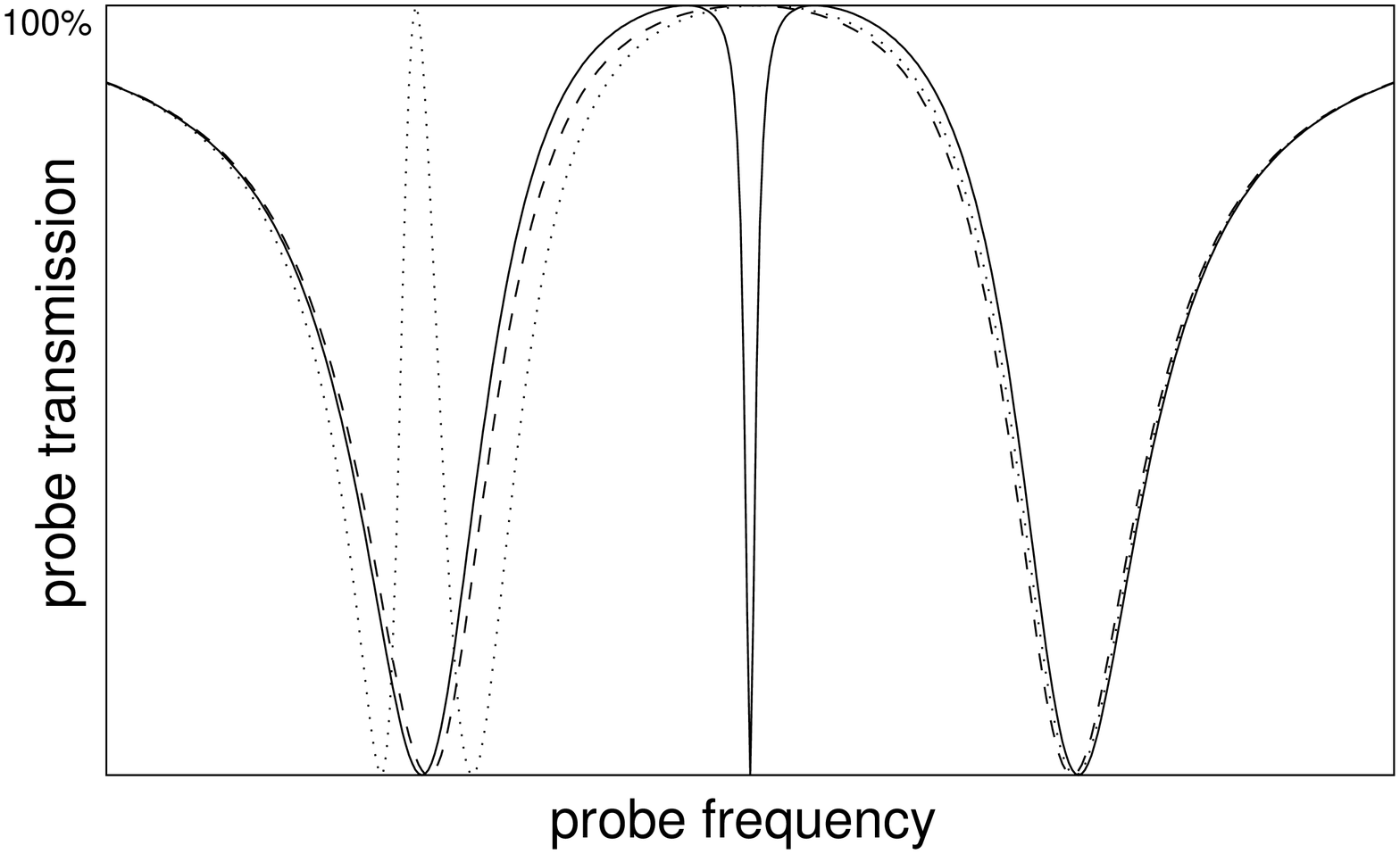}}
 \caption{\protect\label{f_eit}
 Electromagnetically induced transparency on resonance as displayed by
 the $\Lambda$-system in
 Fig.~\protect\ref{f_levels}a (dashed line). The fourth coherently coupled
 level splits the transparency into two and a sharp absorption
 line on resonance appears (solid line). The dotted line is for detuned 
 IR field.
}
\end{figure}

The absorption spectrum of a weak probe field ($\alpha$) can be
changed by coherently preparing a so-called
$\Lambda$-system (states $\left| a \right\rangle$, $\left| b \right\rangle$, and $\left| c \right\rangle$ of
Fig.~\ref{f_levels}). This can be accomplished by a 
strong coherent field (with Rabi frequency $\Omega$) that
gives rise to two interfering Stark split absorption lines \cite{EIT}. When a
fourth state $\left| d \right\rangle$ is coupled by a weak field with Rabi frequency
$\Omega_{\rm IR}$ (from now on referred to as ``IR field'') the resulting
interaction 
Hamiltonian reads 
\begin{eqnarray}
H &=& \hbar\Omega\left|c\right\rangle\left\langle  a\right| +
\hbar\alpha\left| b\right\rangle\left\langle a\right| +
\hbar\Omega_{\rm IR}\left| c\right\rangle\left\langle d\right| +
h.c.\;.
\end{eqnarray}
Without IR field, the ``dark'' state $\left| -\right\rangle = (\Omega\left| b\right\rangle -
\alpha\left| c\right\rangle)/\sqrt{\Omega^2+\alpha^2}$ is decoupled from the optical
fields ($H\left| -\right\rangle = 0$). When the system is driven into this state,
the pair of fields  
propagate through the medium unhindered, i.e. the medium is
transparent on 
resonance (broken line in Fig.~\ref{f_eit}). In case of $\alpha \ll
\Omega$, $\left| b\right\rangle$ nearly coincides with the dark state, thus
basically all population collects in ground state $\left| b\right\rangle$.

A perturbation of the dark state by a coherent field $\Omega_{\rm IR}$
does not necessarily lead to the destruction of coherence. However,
$\Omega_{\rm IR}$ 
can dramatically affect the absorption of the weak probe
field. The transparency splits in two, and results in a ``double dark'' resonance, 
interfering into a very sharp, coherent absorption line in between. Nearly all
electrons stay in ground state $\left| b\right\rangle$.

The propagation dynamics of the probe field is described by the
susceptibility 
\begin{eqnarray}
\chi &=& {\rm i}\eta\frac{\Gamma_{cb}\Gamma_{db} + \Omega_{\rm IR}^2}{\Gamma_{ab}\left(\Gamma_{cb}\Gamma_{db} +
 \Omega_{\rm IR}^2\right) + \Omega^2\Gamma_{db}}\;,
\end{eqnarray}
where $\eta=3\gamma_{a\rightarrow b}{\cal N}\lambda^3/(8\pi^2)$,
 ${\cal N}$ is the electron density, $\gamma_{ij}$ are the relaxation
 rates of the respective coherences and $\Gamma_{ab}=\gamma_{ab} + {\rm i}\Delta$,
 $\Gamma_{cb}=\gamma_{cb} + {\rm i}(\Delta-\Delta_0)$, and
 $\Gamma_{db}=\gamma_{db}+{\rm i}(\Delta-\Delta_0-\Delta_{\rm IR})$, where $\Delta$ ($\Delta_0,
 \Delta_{\rm IR}$) is the detuning of the probe (coupling, IR) field
 to its respective transition. The absorption spectrum of the
 probe field is shown in Fig.~\ref{f_eit}, for the IR field
on (solid line) and off (broken line)\cite{lukin99}. If the
Rabi frequency $\Omega_{\rm IR}$ of the perturbation is weak the
sharp additional 
absorption line of the probe field has approximately Lorentzian
line shape with a width of $
\Gamma \approx \gamma_{a\rightarrow b} \Omega_{\rm IR}^2/\Omega^2 +
\Delta\nu_{\rm IR}$, where $\Delta\nu_{\rm IR}$ is the linewidth of the IR
 field, and at a frequency of $\nu_{\rm new} \approx \nu_0 +
 \Delta_{\rm IR}$, where $\nu_0$ is the 
probe resonance frequency, and $\gamma_{a\rightarrow b}$ the
decay from
state $\left| a \right\rangle$ to $\left| b \right\rangle$. Thus frequency and width 
of this new resonance can be manipulated by changing detuning and
intensity of the IR field. 

In a system where all three fields are on resonance and the
IR Rabi frequency $\Omega_{\rm IR}$ is very small compared to $\Omega$,
the $\Lambda$-type absorption profile is nearly undisturbed everywhere
except for the resonance region. But where there was transparency before
is now a sharp absorption line which splits the transparency in
two (see Fig.~\ref{f_eit}) \cite{lukin99}. 

Note that if  $\Delta_{\rm IR} \approx \Omega$ the sharp, new transparency
resonance appears near the maximum absorption of the $\Lambda$-system
(see Fig.~\ref{f_eit}, dotted 
line). Thus it is also possible to turn absorption 
into transparency.

These features make the present system suitable for a novel kind of detector. In the case of $\Delta_{\rm IR}=0$, i.e. when all three fields
are on
resonance, either both the probe and the IR field are 
absorbed, or none \cite{harris98}. If the IR
field is a long-wavelength infrared field, and the probe
field an optical field, the absorption characteristics of the
visible probe field on resonance translate into  the absorption characteristics of
the IR field. However, the scheme can be used in the opposite mode
as well: If one detunes the perturbation to $\Delta_{\rm IR} \approx \Omega$,
the Stark shifted probe field absorption turns into transmission when
the IR field is present. In this
case, only one of the 
fields is absorbed, either the optical probe field or the IR field,
but never both. 
In the following
paragraphs only the first setup is closely examined. However, both
setups result in  similar sensitivity, and there might 
be cases where the second mode might be more practical \cite{yelin01}.
This ``frequency translation'' can in an obvious way be used not only
for detection but also, e.g., for switching, converting, imaging.

It should be
mentioned here, that for an ideal setup, i.e., where all
three states $\left| b \right\rangle$, $\left| c \right\rangle$, and $\left| d \right\rangle$ are stable and all fields are
monochromatic, this scheme possesses unlimited sensitivity. In any
realistic situation the sensitivity depends on the total ratio
of the decoherence, that is, the lifetime of
the metastable states, and additional incoherent mechanisms such as
phonon scattering, to the strength of the coherent coupling
mechanisms.

The sensitivity for such a system can be found through an
operator-/C-number Langevin approach, assuming $\delta$-correlated
noise. The strength of the
signal is given by the reduction in the probe intensity transmitted when the IR field is turned on:
\begin{eqnarray}
I_{\rm signal} &=&
-\frac{\partial}{\partial \Omega_{\rm IR}^2} \left. I_{\rm total}\rule[-5mm]{0mm}{10mm}
\right\vert_{\Omega_{\rm IR}=0} \Omega_{\rm IR}^2\;.
\end{eqnarray}
With that the efficiency reads
\begin{eqnarray}
\frac{I_{\rm signal}}{I_{\rm IR}} &\simeq&
\underbrace{\frac{\lambda_{\rm
IR}^3}{\lambda_{\rm probe}^3} \,\frac{\gamma_{\rm IR}^{\rm rad}}{\gamma_{\rm probe}^{\rm rad}}}_{\simeq 5}\,\underbrace{
\frac{\alpha^2}{\Gamma\,\gamma_{\rm decoh}}\rule[-4.6mm]{0mm}{10mm}}_{>100}\;,
\end{eqnarray}
where $\gamma_{\rm probe}^{\rm rad}$ ($\gamma_{\rm IR}^{\rm rad}$) is
the radiative decay along the 
probe (IR) transition, $\Gamma$=1-10meV, the decoherence $\gamma_{\rm decoh}$=1meV 
on the most critical transition (in this case the transition between
states $\left| b\right\rangle$ and $\left| c\right\rangle$ in Fig.~\ref{f_levels}a). It turns out that for intraband quantum well systems, the ratio $\lambda_{\rm IR}^2 / \lambda_{\rm probe}^2 \cdot \gamma_{\rm IR}^{\rm rad}/\gamma_{\rm probe}^{\rm rad}$ is usually of the order of one. For the Rabi frequency of the probe field $\alpha\le\Omega\simeq 40$meV, i.e., the saturation point, we can reach an efficiency of up to three orders of magnitude between the signal and the IR field. From equating signal to noise, obtained from a somewhat
lengthy calculation (see \cite{yelin01}) we find the minimum detectable
power of
\begin{eqnarray}
\label{sens_abs}
{\rm P}_{\rm IR}^{\rm min} &\ge& \frac{\hbar\nu_{\rm IR} \,\Gamma}{\sqrt{ \gamma_{\rm
probe}^{\rm rad}t_m}} \, \frac{\gamma_{\rm decoh}}{\Omega} \, \frac{\lambda_{\rm
probe}}{\lambda_{\rm IR}} \, \frac{\gamma_{\rm probe}^{\rm
rad}}{\gamma_{\rm IR}^{\rm rad}} \;\simeq\; \nonumber \\
&& \hbar\nu_{\rm IR}
\frac{\Gamma}{\sqrt{\gamma_{\rm probe}^{\rm rad} t_m}} \,
\frac{\gamma_{\rm decoh}}{\Omega} \, \frac{\lambda_{\rm
IR}}{\lambda_{\rm probe}} \;,
\end{eqnarray}
where $\hbar\nu_{\rm IR}$ is the photon energy of
the IR 
field and $t_m$ the measuring time. For the parameters of the semiconductor example discussed later, with $\lambda_{\rm IR}=10\mu$m, $\Delta\nu_{\rm IR}= 10$GHz this gives the order of 1$\mu$J/sec for a measuring time of one second. Unity optical density is assumed.

Note that complete transparency is ideal but by no means necessary in
order for the detector to perform well: The figure of merit is the
factor $\Omega/\gamma_{\rm decoh}$, which describes effectively the
coherence-to-incoherence ratio in the system, as mentioned above.


For the solid state realization  in
semiconductor quantum well systems, eigenstates can be treated in 
many aspects like atomic states. That is, in Fig.~\ref{f_levels}b the
eigenstates of the uncoupled wells would be analogous to the respective
states in Fig.~\ref{f_levels}a. 
However, in a double well potential $V(z)$ the electrons tunnel
through the very thin barrier between the two wells; thus, the
states (e.g. $\left| a\right\rangle$ and $\left| c\right\rangle$) mix, split, and are shifted
by $\langle a|V(z)\left| c\right\rangle$. Approximately, superposition states of the
uncoupled well states $\left| \pm\right\rangle \propto \left| a\right\rangle\pm\left| c\right\rangle$
emerge. The splitting can be compared with the Stark splitting in
atomic states caused by a strong coupling laser field, $\Omega$, like
in Fig.~\ref{f_levels}a. The two
resulting resonances (corresponding to states $\left| +\right\rangle$ and$ \left| -\right\rangle$)
also interfere 
destructively, so that an EIT-like spectral pattern emerges. In this
case, however, resonant tunneling and not an external monochromatic
field is the coherence generating mechanism. In this case the EIT
phenomenon can be viewed as resulting from Fano-type interference.
We also note that another degree of freedom can be added to this
system (Fig.~\ref{f_levels}b): If one side
(in our case the side of increasing $z$) is ``opened,'' i.e. the
potential of the right side is lowered, a quasi-continuum of states
instead of the discreet eigenstates is found. The magnitude of the
Fano term is directly proportional to the square root of the multiplied
widths of the excited states $\left| +\right\rangle$ and $\left| -
\right\rangle$. This term therefore depends strongly on the excited
state lifetimes.

The theoretical approach to describing states as well as coupling and
dephasing strengths in quantum wells is simple, but gives only an
estimate of the order of 10-50\% accuracy in energy and coupling
strengths. The envelope function of any particular 1-D well geometry
can be determined via a transition matrix method in the case of
discrete states and borrowing the method of calculating Feshbach
resonances (see, e.g., \cite{friedrich}) for quasi-continuous states,
where also the emerging finite lifetimes are found. (These states
tunnel into the continuum, like the one in Fig.~\ref{f_levels}b far right,
above 600meV.) The dipole element between any two states using
envelope functions 
is found via $d=e\left\langle f\right|  z\left| i\right\rangle =
e\int_{-\infty}^{\infty}dz \,
\psi_f(z)\, z \, \psi_i(z)$, where $e$ is the electron charge, and
$\psi_{i/f}(z)$ the 1-D envelope functions. All emerging parameters
can now be plugged into the usual Maxwell-Bloch equations.

The most important phase destroying mechanisms in semiconductor
quantum wells include phonons, non-parabolicity, many-body effects,
and geometrical imperfections. 

The problem we are dealing with is adiabatic and includes no (or
nearly no) excited state population. Thus many-body effects can be
neglected (see, e.g. \cite{dmitri}). With electrons always having
enough time to relax to the bottom of the band before any light
induced transitions,
non-parabolicity can be neglected as well compared to most other
dephasing mechanisms, in particular in GaAs based structures, where the
effect of non-parabolicity is very small.

Phonons are strongly frequency and temperature dependent: Since the
energy gap between state $\left| b\right\rangle$ and the higher lying states is
several hundred meV, not even polar optical phonons are
playing a big  role in population transfer, as it would be the case in
a traditional QWIP, where the electrons have to be transported only
over  a range of few tens of meV.  On the other hand, phonons do play
a role as dephasing agents. The relevant quantity for both optical and
acoustical phonons is the matrix element $G_{if}(q) = \left\langle f\right| {\rm
e}^{\rm iqz}\left| i\right\rangle$, where acoustical phonons are proportional to
$\int_{-\infty}^{\infty} dq \left| G_{if}(q)\right|^2$, and
polar  optical phonons are proportional to
$\int_{-\infty}^{\infty}dq \left| G_{if}
\right|^2/(q^2+Q^2)$, where $Q$ is the in-plane momentum. The upper limits of
acoustical and polar optical phonons ({\it cf.} \cite{greg}) in this proposed structure are
smaller than $10^{-4}$meV and 0.1meV, respectively. 

It thus turns out that the most threatening dephasing mechanism
derives from geometrical imperfections, that is, interface roughness
scattering: The quantum wells and barriers needed for this kind of device are
often only several atomic layers thick. Even in the best molecular
beam epitaxy machines that are found today it is impossible to grow
the required quantum structures completely smooth. But differences in
thickness in the barriers or wells have a relatively strong influence
on the energy of and resonances between
eigenstates. In the experiments on this subject \cite{faist} the
coherence decay due to interface roughness is roughly 0.5-1meV.

The rule of thumb, that most  incoherent broadening can be somehow
 mitigated, probably also applies here: Using the spin components of
 the electronic (or hole) eigenstates should improve the decoherence
 times considerably, in particular if the two states where decoherence
 hurts most (tates $\left| b\right\rangle$ and
 $\left| c\right\rangle$ in
 Fig.~\ref{f_levels}), are spin components of the same electronic eigenstate
 (for details see \cite{yelin01}). 

\begin{figure}[ht]
\epsfxsize=7cm
\centerline{\epsffile{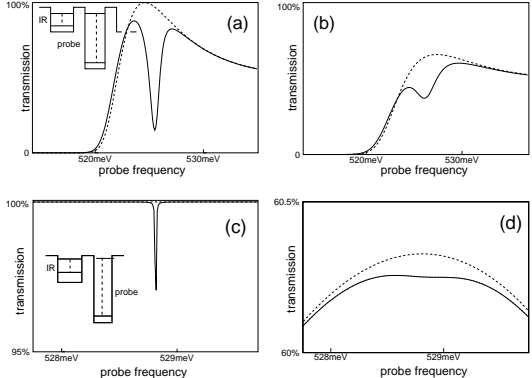}}
 \caption{\protect\label{f_sim}
Transmission spectra for the probe field  with (solid line) and
without (broken line) weak IR field present, simulated for a
GaAs/AlGaAs system. Curves (a,c) are for the ideal case
where no additional dephasing is present, in curves (b,d) dephasing
(phonons, interface roughness scattering, room temperature) is added.
The difference of the upper (a,b) and lower (c,d) systems (see insets)
is the presence/absence of a quasi-continuum for large $z$. The input
IR intensities are 
2.5mW for (a,b), and 500nW for (c,d). Note the
difference in scale for the two examples.}
\end{figure}

The simulations shown in Fig.~\ref{f_sim} have the parametric values
incorporated: Two samples, both without and with additional
(i.e. geometric etc.) dephasing, are tested. Both are
GaAs/Al$_x$Ga$_{1-x}$As structures with high-$x$ wells for a high
offset (800meV) and medium $x$ for the shallow well.  The
emerging offsets and  well/barrier widths are as shown in
Fig.~\ref{f_levels}b. The system is modulation-n-doped to provide the
necessary ground state population. (Optimal electronic density is
assumed which would in practice depend on the number of repetitions of
the structure, beam widths etc.) For the simulations GaAs/AlGaAs
effective masses of 0.067 electron masses are assumed.

Figs.~\ref{f_sim}a,b show a structure with short lived excited states
(notice the ``open'' geometry in the inset) and thus large Fano
factors, (b) has the  dephasing included. The input IR frequency has a
width of $\sim$10GHz (for all Figs.) and an intensity of 2.5 mW/(10$\mu$m)$^2$, with a
parametric dephasing of 1meV. In Figs.~\ref{f_sim}c,d the excited states
are narrow (no quasi-continuum), and the IR power is 0.1$\mu$W/(10$\mu$m)$^2$, with a
parametric dephasing of 0.5meV. (This value seems attainable with
present day technology, {\it cf}. \cite{faist}). 

It is  obvious that this kind of  detector is not broadband, like QWIPs
with the excited state in the continuum, but rather energy selective. For
best results, a frequency filter should be used. On the other
hand, comparing the sensitivity estimate (Eq.~\ref{sens_abs}) with an analogous
one for a QWIP with  comparable coupling strength (which should be only possible with the more restrictive bound-to-bound
state QWIPs) we find for the ratio of the minimum detectable powers:
\begin{eqnarray}
{\rm P}^{\rm coh}_{\rm IR} &=& 
\underbrace{ \frac{\Gamma^{\rm coh}}{\Gamma^{\rm QWIP}}
\rule[-5mm]{0mm}{10mm}}_{\simeq 1} \,
\underbrace{ \frac{\lambda_{\rm probe}}{\lambda_{\rm  IR}}
\rule[-5mm]{0mm}{1mm}}_{\simeq 0.2} \,
\underbrace{ \frac{\sqrt{\gamma_{\rm decoh} \gamma_{\rm probe}^{\rm rad}}}{\Omega}
\rule[-5mm]{0mm}{1mm}}_{\simeq 0.002-0.02} \,
\underbrace{ \sqrt{\frac{\gamma_{\rm decoh}}{\gamma_{\rm decoh}^{\rm QWIP}}}
\rule[-5mm]{0mm}{1mm}}_{1-3}\, 
{\rm 
P}^{\rm QWIP}_{\rm IR}\;,
\end{eqnarray}
where the decoherence in the QWIP results from the broadening of the
lower QWIP state (through phonons, tunneling into continuum).

In conclusion we have demonstrated an example of a new kind of
coherence based nonlinear optical process in semiconductor quantum
wells. Specifically we have shown how it is possible to use this
technique for sensitive photo detection.

We also note that there exists a number of avenues for improvement.
A better frequency 
range for the probe field, e.g. visible or 1.5$\mu$m, can become
accessible by moving the ground 
state into the valence band. In this case, doping would not be
necessary. Further improvement in terms of
coherence lifetimes is expected from utilizing the electronic spin states in
the conduction band. In this case static or dynamic magnetic fields
can be coupled and detected in intraband transitions in the THz range, or polarized
electromagnetic fields in interband transitions. The relevant
coherence lifetimes are expected to be up to four orders of magnitude
higher than for present systems. 

We want to thank L. Friedman, M. D. Lukin, R. Soref, G. Sun
for stimulating and helpful discussions. SFY would like to thank the
Humboldt Foundation for their support.


\vspace{-.4cm}

\begin{thebibliography}{99}

\vspace{-.8cm}

\bibitem{EIT} K.-J.\ Boller, A.~Imamo{\u g}lu, S.~Harris,
Phys.~Rev.~Lett.~{\bf 64}, 2593 (1991); for review see S.~Harris,
Physics Today {\bf 50}, 7, 36 (1997).

\bibitem{hakuta91} K.~Hakuta, L.~Marmet, B.~P.~Stoicheff,
Phys.~Rev.~Lett.~{\bf 66}, 596 (1991)

\bibitem{hemmer95} P.~R.~Hemmer {\it et al.}, Opt.~Lett.~{\bf 20}, 982
(1995); T.~T.~Grove {\it et al.}, Opt.~Lett.~{\bf 22}, 769 (1997);
T.~T.~Grove {\it et al.}, Opt.~Lett.~{\bf 22}, 1677 (1997).

\bibitem{lukin97} M.~D.~Lukin {\it et al.}, Phys.~Rev.~Lett.~{\bf 79},
2959 (1997).

\bibitem{hakuta97} K.~Hakuta {\it et al.}, Phys.~Rev.~Lett.~{\bf 79},
209 (1997).

\bibitem{faist} J.~Faist et al., Nature {\bf 390}, 589 (1997); A.~Imamo{\u g}lu,
R.~J.~Ram, Opt.~Lett.~{\bf 64}, 873 (1994); H.~Schmidt et al.,
Appl.~Phys.~Lett.~{\bf 70}, 3455 (1997). 

\bibitem{sera00} In G.~B.~Serapiglia et al., Phys.~Rev.~Lett.~{\bf
84}, 1019 (2000) the transparency is not created via resonant tunneling.

\bibitem{dmitri} D.~E.~Nikonov, A.~Imamoglu, L.~V.~Butov,
Phys.~Rev.~Lett.~{\bf 79}, 4633 (1997); D.~E.~Nikonov, A.~Imamoglu,
M.~O.~Scully, Phys.~Rev.~B {\bf 59}, 12212 (1999).

\bibitem{lukin99} M.~D.~Lukin {\it et al.}, Phys.~Rev.~A {\bf 60},
3225 (1999).

\bibitem{harris98} S.~Harris, Y.~Yamamoto, Phys.~Rev.~Lett. {\bf 81},
3611 (1998).

\bibitem{yelin01} S.~F.~Yelin and P.~R.~Hemmer, ``Sensitivity
estimates for a coherence based detector in semiconductor nanostructures'' (in preparation).

\bibitem{friedrich} H.~Friedrich, ``Theoretical Atomic Physics,'' Springer, Berlin (1998).

\bibitem{greg} G.~Sun, Y.~Lu, L.~Friedman, R.~A.~Soref, Phys.~Rev.~B {\bf 57}, 6550 (1998); G.~Sun, L.~Friedman, Phys.~Rev.~B {\bf 53}, 3966 (1996).

\end{thebibliography}
\end{document}